\documentstyle[epsf]{myl-aa}

\begin {document}
   \thesaurus{03 (09.03.2; % cosmic rays
              02.13.1; % Magnetic fields
               11.09.3; % intergalactic medium
               13.07.2) % gamma rays, observation
              }
   \title{An extragalactic ``flux trapping'' origin of the dominant
part of hadronic cosmic rays?}
\author{R.Plaga 
%\inst{1}
%\and  S.M. Bradbury \inst{1}
}
\institute{Max-Planck-Institut f\"ur Physik, F\"ohringer Ring 6,
        D-80805 M\"unchen, Germany
%\and Max-Planck-Institut f\"ur Kernphysik, P.O. Box 103980,
%        D-69029 Heidelberg, Germany
}
   \offprints{plaga@hegra1.mppmu.mpg.de}

   \date{Received 14 August 1997 ; accepted 28 October 1997}

   \maketitle

\vspace{-0.3cm}

\begin{abstract}
An extragalactic origin of the dominant part
of all extrasolar hadronic cosmic rays 
above about 10 MeV/nucleon has long
been considered unlikely due to energy
considerations. 
In order to circumvent such arguments,
the hypothesis that ``flux trapping'' of extragalactic
cosmic rays occurs in the Galactic confinement volume is
advanced in this paper. This hypothesis
is based on a number of speculative
assumptions about the properties of Galactic and
intergalactic magnetic fields.
%The most speculative 
%assumptions for this hypothesis are:
%\\
%1. that
%major parts of the intergalactic magnetic field
%surrounding the Galaxy connect to the Galactic field, and 
%\\
%2. the conservation the adiabatic invariant in intergalactic
%cosmic-ray propagation. 
%\\
%Liouville's theorem then leads to the expectation of an
%enhanced density of extragalactic cosmic rays
%in the Galactic confinement volume. 
The intergalactic cosmic-ray density 
expected under conservative assumptions about its
extragalactic origins
is then shown to be of the right order
of magnitude to account for the locally observed
cosmic radiation. It is demonstrated that
an extragalactic scenario of cosmic-ray origin
that is consistent with the observed
cosmic-ray energy spectrum and preserves
the successes of Galactic propagation theory
can be constructed.
The position of the ``ankle'' in the cosmic-ray
energy spectrum follows as a natural consequence
from this explanation. 
The $\gamma$-ray flux from the Magellanic clouds
is shown to provide no suitable testing
ground for the decision for or against
an extragalactical origin in this scenario. 
It is argued that recent observational evidence
seems to be in favour of a dual origin of cosmic-rays.
The hadronic component is mainly extragalactic
while the electrons are accelerated in Galactic supernova remnants. 

\vspace{-0.3cm}
\vspace{-0.5cm}
\end{abstract}
\vspace{-0.5cm}
\section{ Introduction }
The possibility of an extragalactic origin
of the dominant part of extrasolar hadronic cosmic rays
with energies above about 10 MeV/nucleon
has been considered since the early days
of cosmic-ray research (Baade $\&$ Zwicky \cite{baade}). 
In the modern era it was discussed by Burbidge(\cite{burbidge}), 
Burbidge $\&$ Hoyle(\cite{burbidge2}) and
later in great detail by Brecher $\&$ Burbidge(\cite{brecher}).
Their ideas were criticized by Ginzburg(\cite{ginzburg})
and others, mainly on energetic grounds: it was argued
that under reasonable assumptions the
potential extragalactic sources of hadronic cosmic rays
like radio galaxies, active galactic nuclei and normal galaxies are
expected to fill the universe with a cosmic-ray energy density
$\rho_{\mathrm{eg}}$ of
about 10$^{-4}$-10$^{-5}$ eV/cm${^3}$
during its lifetime t$_{\mathrm{U}}$ $\simeq$
1.5 $\cdot$ 10$^{10}$ years,
rather than the locally observed value 
$\rho_{\mathrm{loc}}$ $\simeq$
0.5 eV/cm$^3$. Section \ref{flux} presents a hypothetical
explanation for the local cosmic-ray density which avoids
the above conclusion by way of a ``density enhancement''
in the Galaxy. The present knowledge about intergalactic
cosmic-rays is discussed in connection
with this explanation in Sect.\ref{origin}. Sect.
\ref{scenario} contains a speculative scenario for the origin
of the observed cosmic-ray energy spectrum.  
\\
There is now impressive 
experimental evidence for an origin of the electron part
of the cosmic rays in Galactic supernova remnants (SNRs)
(see Sect.\ref{origin}). 
On the other hand
no direct, incontrovertible
evidence for any Galactic source of hadronic
cosmic rays has been found up to now.
An extragalactic origin for hadronic cosmic
rays below about 10$^{18}$ eV has therefore 
generally been considered 
unlikely but not impossible(Longair \cite{longair}). 
After a discussion of the observational evidence in
Sect.\ref{observation},
I will argue in the conclusion that an extragalactic origin
of the hadronic and a Galactic origin of the electron component
could be the most natural explanation of the observational
facts.
A few years ago Sreekumar et al.(\cite{sreekumar}) interpreted
their upper limit
on $\gamma$-radiation with energies above 100 MeV
from the Small Magellanic cloud (SMC) 
obtained by the EGRET detector
as excluding an extragalactic origin of cosmic rays.
Section \ref{smc} examines this conclusion in the light of
ideas of this paper.
\\
Recently very intense VHE $\gamma$ radiation
(energy range 0.3 -10 TeV)
of extragalactic origin was observed
by several groups (see e.g. Deckers et al. \cite{deckers}).
These observations may well turn out to be
the first direct experimental evidence
for an acceleration site of
hadronic cosmic rays
(``the proton blazar'', see Mannheim et al. (\cite{mannheim})). 
It  was this unexpected discovery that
prompted the present author to
reconsider the ``extragalactic option'' for
the origin of the main part of hadronic cosmic
rays. 
\vspace{-0.5cm}
\section{Necessary conditions for an extragalactic origin 
of cosmic rays and
intergalactic diffusion}
\label{diffsec}
Before I discuss the specific assumptions for
the hypothesis of this paper I list the basic
conditions which have to be fulfilled for
an extragalactic origin of cosmic rays.
Low energy cosmic rays can reach earth from
intergalactic space if the following conditions are
fulfilled:

\noindent$\bullet$ a hypothetical Galactic wind
streams outwards from the Galactic centre
with a velocity below about 30 km/sec
so that cosmic-ray transport is dominated by
diffusion rather than by
convection. This may well be true 
because it is uncertain if a universal Galactic wind
exists at all (Berezinskii et al. \cite{berez}).
There is some evidence that our Galactic disk
may have ``chimneys'' (Norman $\&$ Ikeuchi \cite{norman2}) 
where interstellar matter is
rapidly streaming out, but {\it most} of the disk's
volume may satisfy the mentioned limit on velocity.
\\
\noindent$\bullet$ the field lines of the galactic-halo
magnetic field are not entirely closed, therefore
cosmic-ray particles of all energies can enter the Galaxy.
This point is discussed in detail in the next section.
%Field-line crossing is thought to occur 
%during the cosmic-ray propagation in the Galactic disk,
%I therefore think it is very unlikely that the Galaxy 
%is magnetically ``hermetically sealed''.
%This general opinion is shared by Ginzburg\cite{ginzbook},
%who argues with the experiences in the confinement 
%of laboratory plasmas.
\\
\noindent $\bullet$ the diffusion coeffcient D$_{\mathrm{IG}}$,
which describes the intergalactic propagation 
of cosmic rays near our Galaxy (i.e. 
outside dense galaxy clusters and galaxy poor ``cosmological
voids'')
has to be large enough
to allow the diffusion of cosmic-ray particles from
extragalactic cosmic-ray sources located
at a distance d from 
our Galaxy within the lifetime of the universe t$_{\mathrm{U}}$:
\begin{equation}
D_{\mathrm{IG}} > {(d/4 Mpc)^2 \over (t_{\mathrm{U}}/1.5
\cdot 10^{10} years)} \simeq
3 \cdot 10^{32} cm^2 sec^{-1}
\label{DIG}
\end{equation}
Here 4 Mpc is the distance to Cen A, a nearby
very extended radio galaxy which had been identified
by Burbidge(\cite{burbidge}) 
as a likely source of extragalactic cosmic rays.
The actual value of D$_{\mathrm{IG}}$ is unknown presently.
Ginzburg $\&$ Syrovatzkii (\cite{ginz}) give
limits\footnote{
I converted these limits to an
intergalactic matter density of 10$^{-6}$/cm$^3$ instead of
10$^{-5}$/cm$^3$ as assumed by
Ginzburg $\&$ Syrovatzkii. Even lower numbers would raise 
the lower limit.} on D$_{\mathrm{IG}}$ of  10$^{31}$ cm$^2$ sec$^{-1}$ $<$
 D$_{\mathrm{IG}}$ $<$  10$^{35}$ cm$^2$ sec$^{-1}$.

The strength of the intergalactic magnetic 
field B$_{\mathrm{IGM}}$ near our Galaxy is unknown.
It could be extremely low 
(below 10$^{-18}$ G) if galactic winds and jets
from active objects did not manage to fill
the universe with magnetized plasma (Kronberg \cite{kronberg}).
Otherwise a mean value very roughly around 
10$^{-11}$ G has been estimated by Daly $\&$ Loeb (\cite{daly}).  
I estimated a modern estimate for  D$_{\mathrm{IG}}$ 
by scaling the 
most likely value of the intracluster 
diffusion coefficient D$_{\mathrm{IC}}$ 
in the Coma cluster
as derived by Schlickeiser et al.(\cite{schlickeiser}) 
(D$_{\mathrm{IC}}$ $\simeq$ 10$^{29}$  cm$^2$ sec$^{-1}$)
for an assumed intracluster magnetic field of
3 $\cdot$ 10$^{-6}$ G, to an intergalactic field
with field strength of 10$^{-11}$ G.
For this I used the assumption, made recently 
by V\"olk et al.(\cite{voelk}),
that D $\sim$ E$^{0.5}$ in the intergalactic medium.
One derives an estimate of D$_{\mathrm{IG}}$ $\simeq$ 10$^{32}$ 
cm$^2$ sec$^{-1}$. If intergalactic fields are weaker
the number could be higher.

In summary it seems quite likely that inequality
(\ref{DIG}) is in fact fulfilled but possibly not 
by a large margin.
%\begin{enumerate}

%\item{    
%      Calibration and flat fielding are based on regular measurements
%      of the pedestals and the relative photomultiplier gains.}
%\item{In addition to a 
%hardware trigger condition of 2 out of 61 pixels fired (including at least 1 of the 
%central 37) a software trigger 
%condition of 2 out of 37 were used to exclude camera-edge events with incomplete images.
%A CONC$<$0.95 cut and data from an
%anticoincidence shield was used to reject events due to cosmic ray muons. Events recorded
%under poor telescope positioning were rejected leaving a mean absolute pointing error of 0.1$^\circ$
%for events after filter.
%    Each observation was scanned for short-term rates significantly above or below the mean rate of that
%    run and run pairs containing abnormal peaks were removed.
%\footnote{The method used here is similar to that of Reynolds et al. (\cite{reynolds}) except that here 
%core pixels are required to have at least one adjacent core pixel.} 
%}
%\item{
%In order to enhance the signal, a series of image parameter cuts was applied which enrich the sample
%with $\gamma$-shower candidates. These cuts were
%optimized on Monte Carlo simulations and real data (30/30 hours of ON/OFF observations of the Crab Nebula 
%made with CT2 in 1994 at $\theta$ $<$ 30$^\circ$).
%\begin{center}
%\begin{tabular}{c}
%        $ 0.6^\circ < \mathrm{DIST} < 1.2^\circ$\\ 
%   $ 0.06^\circ < \mathrm{WIDTH} < 0.2^\circ$ \\
%   $ 0.18^\circ < \mathrm{LENGTH} < 0.375^\circ$\\
%   $ 0.5 < \mathrm{CONC} $ \\
%   $ \mathrm{ALPHA} < 15^\circ $ \\
%\end{tabular}
%\end{center}
%}
%\end{enumerate}
\vspace{-0.2cm}
\section{Magnetic flux trapping in the Galaxy}
\label{flux}
%The phenomenon of ``flux trapping'' is well known from
%the technology of high-flux nuclear reactors
%\cite{nucl}: in order to enhance the
%neutron flux, fast neutrons produced via fission
%are directed into a moderator in the middle of the
%reactor core. Because the lifetime
%of neutrons in the moderator is much longer 
%than elsewhere in the core (because of lower neutron
%absorption cross sections) the flux in the moderator 
%can be enhanced over the one in rest of the core 
%by a factor of about a hundred.
In this Sect. the central hypothesis
of this paper that extragalactic cosmic rays
are ``concentrated'' in the Galaxy
is discussed.
Parker (\cite{parker}) has long argued that the magnetic
field lines of the Galaxy are generally closed.
Due to the dynamical pressure of confined cosmic radiation,
instabilities form along the field lines, and the magnetic
field together with relativistic plasma is expelled
out of the Galactic confinement volume. 
These loops are assumed to be the places
where cosmic rays escape from the Galaxy (Jokipii $\&$
Parker \cite{jokipii}).
\begin{figure}
	 \epsfxsize=8.7cm
	 \epsffile{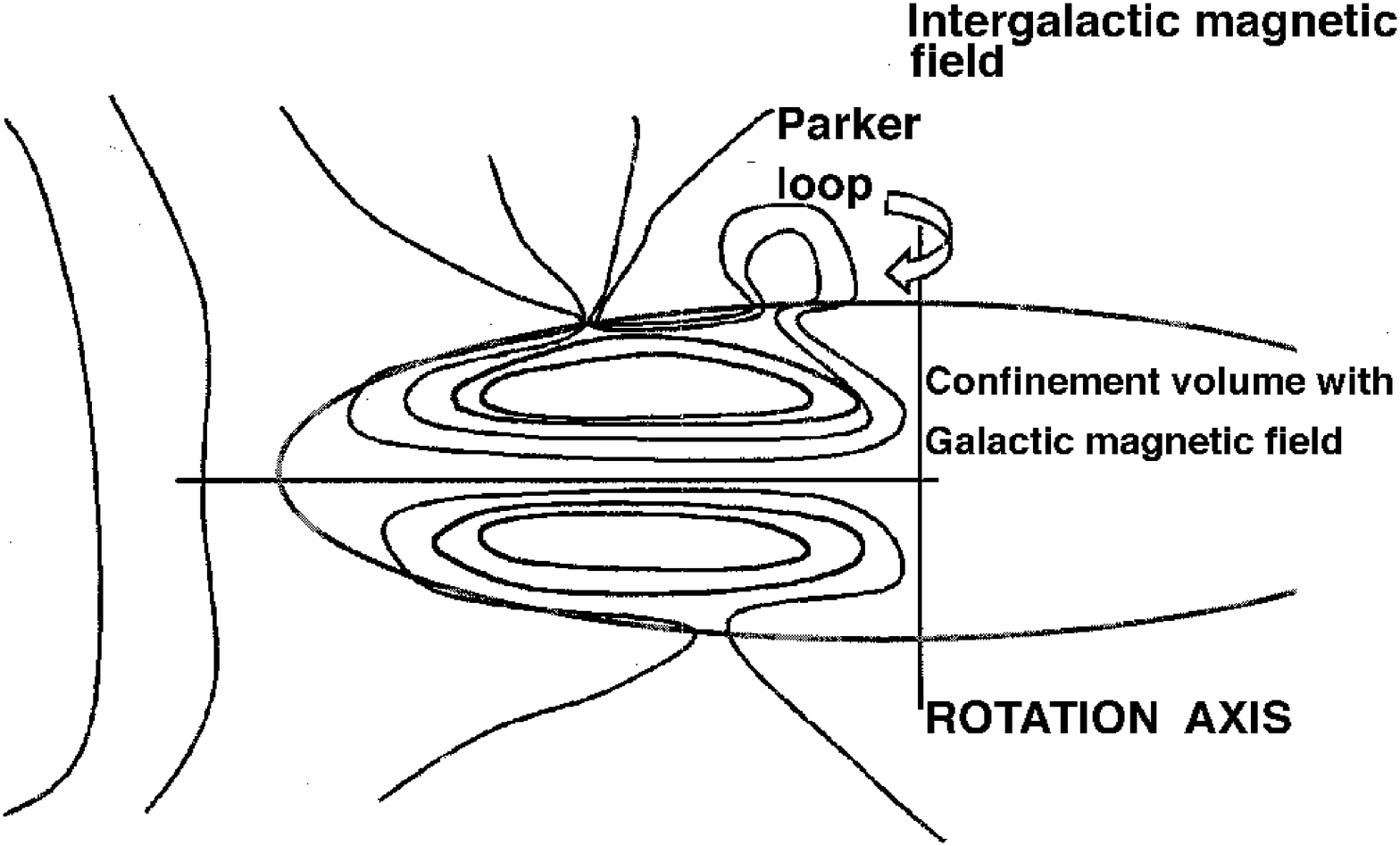}
	\vspace{-0.5cm}
         \caption[]{\small \label{magstruc}
Schematic sketch of the hypothetical general
topology of the Galactic and intergalactic magnetic field
assumed for the present scenario. The field is
generally closed,
but some Parker loops connect to the intergalactic magnetic
field. Cosmic rays enter through the connected loops and
escape through all loops.
 } 
	\vspace{-0.3cm}
\end{figure}
Let us assume that
in a fraction of cases some dissipational mechanism exists
and the magnetic field of these ``Parker loops''
reconnects with the ambient intergalactic field 
(Parker \cite{parker92}). 
In this case the intergalactic cosmic rays on the 
connected field lines could
enter the Galaxy(fig.\ref{magstruc}).  
A ``reconnection factor'' $f_{\mathrm{r}}$ can be defined as:
\begin{equation}
f_{\mathrm{r}} = A_{\mathrm{f}}/A_{\mathrm{g}}
\label{reconf}
\end{equation}
where A$_{\mathrm{f}}$ is the area in intergalactic space
far away from the Galaxy perpendicular
to the field direction from which field lines
connect to the Galaxy, and A$_{\mathrm{g}}$ is the geometrical
area of the Galactic disk.
\\
{\it The assumptions for the hypothesis
of ``flux trapping'' advanced in this paper
can be summarized as:
\\
\noindent 1.  $f_{\mathrm{r}}$ $\simeq$ 1 with a predominantly
but not completely closed Galactic magnetic field geometry.
\footnote{A completely open field geometry,
as would be expected from a primordial origin 
of the Galactic magnetic field, is improbable for reasons
which have been discussed 
in connection
with an extragalactical 
origin of cosmic rays by Parker (\cite{parker73}).}
\\
2. A relatively small intergalactic
field strength (B$_{\mathrm{IGM}}$ $<$ 10$^{-10}$ G, see below).
\\
3. Conservation of the ``adiabatic invariant'' 
sin$^2$($\theta$)/B$_{\mathrm{IGM}}$ (Ginzburg $\&$
Syrovatskii \cite{ginz})
in intergalactic propagation of cosmic rays. Here $\theta$
is the pitch angle. This amounts to the assumptions
that distances with variations in the intergalactic field
are generally large with respect to distances associated
with particle motions. Taking into account that 
e.g. for an energy of 1 GeV and B$_{\mathrm{IGM}}$ $\simeq$
10$^{-11}$ G the gyro-radius is about 1 light year
the assumption seems reasonable. It is however
the most controversial of all listed assumptions.
\\
4. Origin of extragalactic cosmic rays in regions with
a magnetic field strength B$_{\mathrm{er}}$ comparable or
larger than the Galactic field
strength  B$_{\mathrm{Gal}}$ and a metallicity not
very different from the Galactical one.
The former assumption is quite
plausible because galaxies,
active objects and galaxy clusters are known to possess
such field strengths (Kronberg \cite{kronberg}).
The equilibrium field strengths in the lobes of radio
galaxies also usually fulfill this condition (Ginzburg $\&$
Syrovatskii \cite{ginz}). The latter assumption can 
be fulfilled under various circumstances (see sect.
\ref{origin}).}
\\
If cosmic rays escape from a high field region
(assumption 4) into a low field region conserving
the adiabatic invariant (assumption 3)
they have small pitch angles below
a maximal angle
$\theta_{\mathrm{max}}$ $=$ arcsin(
$\sqrt{B_{\mathrm{IGM}}/B_{\mathrm{er}}}$) (Ginzburg $\&$
Syrovatskii \cite{ginz}).
Under these conditions the particles can freely
enter the Galactic confinement volume via the open field lines.
%The density of extragalactic
%cosmic rays upon entry
%in the Galaxy would then 
%be enhanced by a factor 
%b$_{\mathrm{conc}}$ $\simeq$ B$_{\mathrm{Gal}}$/B$_{\mathrm{IGM}}$
%with respect to the value $\rho_{\mathrm{eg}}$
%far away from any galaxy.
\\
Because of various inhomogeneities of the Galactic
magnetic field down to very small scales, the adiabatic
invariant is generally assumed {\it not} to be conserved
during Galactic propagation (Cesarsky \cite{cesarsky}):
The cosmic rays are expected to fill all
Galactic field lines during their propagation 
by spreading to other field lines due to the irregular
component of the Galactic magnetic field(Ptuskin \cite{ptuskin}).
All of momentum space is filled due to pitch-angle scattering, e.g. by
hydromagnetic waves.
\\
Even with f$_{\mathrm{r}}$ $\simeq$ 1 (assumption 1 above)
there could be Parker loops which do not
connect to  intergalactic
field lines.
Let us first consider the limiting case 
in which {\it all} Parker loops
are fully connected, however.
Liouville's theorem would then predict a cosmic-ray
density enhanced by b$_{\mathrm{conc,equ}}$ 
relative to the intergalactic value throughout the
Galaxy with the same energy spectrum as in
intergalactic space. 
This is because according to Liouville's theorem
the directional differential intensity along a
field line has to be constant, while the radiation
is constrained to move with small pitch angles
outside but not inside the Galaxy
due to the conservation of the adiabatic invariant.
Both conditions together can only be fulfilled
with an enhanced total density of cosmic radiation
inside relative to the outside of the Galaxy.
The flux in the weaker field is reduced in proportion
to the permitted cone. This solid angle ``enhancement''
factor is given for the case of 
$B_{\mathrm{Gal}}/B_{\mathrm{IGM}} \ll 1$  (Ginzburg $\&$
Syrovatskii \cite{ginz}):
\begin{equation}
b_{\mathrm{conc,equ}}=2B_{\mathrm{Gal}}/B_{\mathrm{IGM}}
\label{bconc}
\end{equation}
For an assumed intergalactic field
strength of 10$^{-11}$ G and a typical
Galactic field strength B$_{\mathrm{Gal}}$ of a few $\mu$G, 
a density enhancement of more than a factor 10$^5$
would be expected\footnote{This 
concentration mechanism
was already briefly mentioned by Burbidge(\cite{burbidge})
and is discussed in greater detail in a context related
to the present one by Sciama(\cite{sciama}).}.
\\
In reality, the above conclusion of a concentration
factor of b$_{\mathrm{conc,equ}}$ is probably too extreme.
Once the cosmic-ray energy
density rises above the energy density of the Galactic
magnetic field, additional Parker loops are expected to form.
These do not connect to intergalactic field lines,
and there cosmic rays could leave the Galaxy e.g. via
loops ``detached'' from the Galactic magnetic field 
(Parker \cite{parker92}).
The mean lifetime of cosmic rays in the Galaxy would drop,
until an equilibrium between the incoming
extragalactic cosmic rays and the total loss
of cosmic rays from the Galaxy, 
at a total pressure similar to the Galactic
magnetic-field energy density, is reached.
Let us assume that cosmic rays mainly
escape via Parker loops through which they
do not enter. This will be the case if the
cosmic-ray equilibrium density, equal to
the local density at earth $\rho_{\mathrm{loc}}$
\footnote{The local density
is taken to be representative for the one in the 
Galactic confinement volume in general, in this
paper.},  is much smaller than the product
of extragalactical cosmic-ray density $\rho_{\mathrm{eg}}$ and the
concentration factor b$_{\mathrm{conc,eq}}$.
This condition corresponds to intergalactic
field strengths below about 10$^{-10}$ G according
to Eq.(\ref{bconc}) together with the
cosmic-ray energy densities quoted in the introduction.
The local density of cosmic rays $\rho_{\mathrm{loc}}$ will
then be determined
by the total influx of extragalactic cosmic rays (parametrized
by f$_{\mathrm{r}}$ and the extragalactic density $\rho_{\mathrm{eg}}$) and
the confinement time $t_{\mathrm{conf}}$ of cosmic rays in the Galaxy.
This density 
will be enhanced over the intergalactic value
by a factor $e$ which is given as:
\begin{equation}
e \equiv \rho_{\mathrm{loc}} / \rho_{\mathrm{eg}}={f_{\mathrm{r}}} 
{t_{\mathrm{conf}} \over t_{\mathrm{equiv}}}
\label{enhance}
\end{equation}
Here $t_{\mathrm{conf}}$ is the lifetime of cosmic-ray
in the Galactic disk and halo (the confinement volume)
due to magnetic diffusion generated trapping.
 t$_{\mathrm{equiv}}$ is the lifetime of cosmic-ray particles
in an equivalent volume of intergalactic space. 
Equation (\ref{enhance}) is valid only if the cosmic rays
mainly escapes through pathways through which it
does not enter, as in the above scenario.
More generally, only an inequality $e$ $\leq$ 
f$_{\mathrm{r}}$ ${{t_{\mathrm{conf}}} \over {t_{\mathrm{equiv}}}}$ is valid.
In this situation Galactic propagation is expected
to be very similar to the one in present
models which have considerable
experimental support (Ferrando \cite{ferrando}).
Only the ``source'' of hadronic cosmic rays is not some
Galactic object class, but the influx of extragalactic cosmic
rays. 
\\
Let us estimate the enhancement $e$ expected from the
parameters of Galactic cosmic-ray
propagation theory, which are based on experimental data.
The lifetime of particles in the Galactic confinement volume is
given as:
\begin{equation}
t_{\mathrm{conf}} = {d^2 \over D_{\mathrm{G}}}
\label{conf1}
\end{equation}
where d is smallest extension of the confinement volume
(the thickness of the disk if it is assumed
to have a cylindrical shape)
and D$_{\mathrm{G}}$ is the Galactic diffusion coefficient.
I will assume parameter values\footnote{These
values were derived under the assumption
that cosmic rays are produced in the Galaxy.
They remain approximately valid
for our case if the mass density
crossed by the particles
prior to the entry is negligible.
Intergalactic propagation during t$_{\mathrm{U}}$
would only contribute about 10$^{-4}$ to the
total grammage crossed by cosmic-ray
particles.
In using these values I assume that the propagation in
the extragalactic acceleration site 
is negligible compared to the later Galactic one (
see discussion in Sects.
\ref{origin} and \ref{observation}).}  derived in a 
diffusion model based on experimental data with $^{10}$Be
at an energy per nucleus of 3 GeV (Ferrando \cite{ferrando}):
d $\simeq$ 3 kpc, D$_{\mathrm{G}}$  $\simeq$ 10$^{28}$ cm$^2$ sec$^{-1}$
and consequently t$_{\mathrm{conf}}$  $\simeq$ 3 $\cdot$ 10$^8$ years.
The lifetime $t_{\mathrm{equiv}}$ of a particle in an equivalent
volume in intergalactic space, is given by:
\begin{eqnarray}
t_{\mathrm{equiv}} = {d^2 \over D_{\mathrm{IG}}} \ \ \ \ \ \ \ \ \ {for\, D_{\mathrm{IG}}/d < c}
\label{equiv1}
\\
 t_{\mathrm{equiv}} = {d \over c} \ \ \ \ \ \ \ \ \  {for\, D_{\mathrm{\mathrm{IG}}}/d > c}
\label{equiv2}
\end{eqnarray}
Assuming that
Eq.(\ref{DIG}) holds,
Eq.(\ref{equiv2}) has to be used for
the calculation of  $t_{\mathrm{equiv}}$ and
I finally obtain (with f$_{\mathrm{r}}$ =1) :
 \begin{equation}
e = {{d c} \over { D_{\mathrm{G}}}} \simeq 3 \cdot 10^4 
\label{eval}
\end{equation}
\vspace{-0.1cm}
One can compare this with the energy density
of intergalactic cosmic rays of 10$^{-4}$-10$^{-5}$ eV
/cm$^3$ estimated by
critics of the extragalactic hypthesis of cosmic-ray
origin, quoted in the introduction:
Remarkably one gets as a natural and
``untuned'' consequence of our scenario:

$\rho_{\mathrm{loc}}$ $\simeq$ $\rho_{\mathrm{eg}}$ $\cdot$ $e$

The predicted enhancement factor $e$
has the right order of magnitude to explain the
local energy density of cosmic rays, under the assumption 
of an extragalactical cosmic-ray energy density which is
considered likely by most workers in the field.
\\
The above ``flux trapping'' hypothesis has a certain similarity
to the particle trapping in the van Allen radiation belts around earth,
where the low energy 
cosmic-ray density is concentrated by a factor
of about 10$^{4}$ relative to the
interplanetary value due to the action of
a relatively strong magnetic fields on cosmic
rays(Hess \cite{hess2}). In both cases  the general conditions and
physical mechanisms are different, however\footnote{
Another analogous situation is the flux trapping in
high-flux nuclear fission reactors. The neutron density
is increased in the moderator because neutrons crossing 
the moderation zone have a larger lifetime 
due to a smaller capture cross section
in ${^{238}U}$(Byrne \cite{byrne}). Cosmic rays crossing the Galactic 
confinement volume have a larger lifetime than in an
``equivalent volume'' of intergalactic space because of a lower
``effective speed'' due to the action of diffusion.}. 
%Such an effect is also responsible for the high concentration
%of particles in the outer van Allen belt.
\vspace{-0.15cm}
\section{On the density and origin of intergalactic cosmic rays}
\label{origin}
\vspace{-0.15cm}
Under which circumstances do intergalactic cosmic
rays have the right density to account for the
local hadronic cosmic rays under the assumptions
of the previous section, and what is their origin?
The expected energy density of extragalactic cosmic
rays $\rho_{\mathrm{eg}}$  relative to the locally
observed  $\rho_{\mathrm{loc}}$
one, can be roughly estimated to be:
\begin{equation}
\begin{array}{rcl}
\Large
\rho_{\mathrm{eg}}/\rho_{\mathrm{loc}} &=& 10^{-4} \times 
\left(\frac{\Omega_{\mathrm{B}}}{0.01}\right)
\times
\left(\frac{\rho_{\mathrm{c}}}{10^{-29} g}\right) 
\times
\left(\frac{M_{\mathrm{g}}}{2 \cdot 10^{44}g}\right) 
\times  \\ & &
\left(\frac{\epsilon}{3 \cdot 10^{40} erg/sec}\right) 
\times 
\left(\frac{{\rho_{\mathrm{BP}}}/\rho_{\mathrm{N}}}{10}\right) 
\times  \\
& &
\left(\frac{\rho_{\mathrm{AGN}}/\rho_{\mathrm{N}}}{1}\right) 
\times
\left(\frac{\epsilon_{\mathrm{h}} /\epsilon_{\mathrm{e}}}{100}\right) 
\times
\left(\frac{V_{\mathrm{U}}/V_{\mathrm{conf}}}{1}\right) 
\end{array}
\label{int1}
\end{equation}
The symbols are defined as:

$\Omega_{\mathrm{B}}$ baryonic mass fraction relative 
to critical density $\rho_{\mathrm{c}}$

$\rho_{\mathrm{c}}$ is the cosmological critical mass density

$M_{\mathrm{g}}$ is the mass of a galaxy similar to the Milky Way

$\epsilon$ is the cosmic-ray production rate of such a galaxy

{$\rho_{\mathrm{BP}}$,$\rho_{\mathrm{N}}$} are the total amount of cosmic rays
produced by such a galaxy during its early bright starburst phase
and later normal life respectively(V\"olk et al. \cite{voelk})

{$\rho_{\mathrm{AGN}}/\rho_{\mathrm{N}}$ is the total amounts of cosmic rays
produced by active objects (radio galaxies, active nuclei,
quasars etc.) relative to the amount produced by
normal galaxies}. The values for both this and the previous
parameter are extremely uncertain and could be very
different from the ones chosen in Eq.(\ref{int1}) (see discussion
at the end of this section.)

{${\epsilon_{\mathrm{h}} / \epsilon_{\mathrm{e}}}$ is the relative 
efficiency of hadron and electron acceleration}

{$V_{\mathrm{conf}},V_{\mathrm{U}}$ 
is the intergalactic confinement volume
(e.g. the volume of all superclusters, see below)
for hadronic cosmic rays
and the volume of the universe respectively}

The numbers displayed are standard choices, and lead
to a value of  $\rho_{\mathrm{eg}}$ with an order
of magnitude required for an 
extragalactic origin in the scenario 
discussed in Sect.\ref{flux}.
 \\
There is an alternative non-standard possibility to choose
the parameter values, which leads to the same
density $\rho_{\mathrm{eg}}$ and is more likely in my opinion. 
It is well known that electrons
cannot be of extragalactic origin because at
energies above about 1 GeV they loose
energy by Compton scattering too quickly
to travel extragalactic distances (Longair \cite{longair}).
Moreover there is recent  
experimental evidence for a supernova-remnant (SNR)
origin of Galactic cosmic-ray electrons
(Koyama et al. \cite{koyama}; Mastichiadis $\&$ de Jager \cite{masti}).
However, if the locally observed cosmic-ray electrons
are produced by SNRs and ${\epsilon_{\mathrm{h}} / \epsilon_{\mathrm{e}}}$=100,
Galactic and intergalactic hadronic cosmic rays
would have comparable intensities.
I will argue in Sect.\ref{observation} that some observational
facts are explained in the most natural way if hadronic cosmic rays
have a {\it mainly} extragalactical origin.
A plausible parameter choice in Eq.(\ref{int1}) is therefore that 
${\epsilon_{\mathrm{h}} / \epsilon_{\mathrm{e}}}$ is 
roughly on the order of 1
\footnote{Donahue (\cite{donahue}) has argued for such a choice
as the most natural one in the Fermi acceleration
mechanism. He then
concluded from the small observed 
cosmic-ray intensity of electrons relative
to hadrons, that a large part of primary electrons had to be lost
during propagation
due to inverse Compton scattering on photons. This excluded
a Galactic origin due to a resulting
too small photon surface density.
Donahue identified a local solar (with high photon densities)
and an extragalactical origin (with large pathlengths)  
as the remaining possibilites. Although the
microwave background radiation was not known when 
Donahue wrote his paper, his conclusion remains valid when
making his assumption about acceleration efficiency
of hadrons.}
and e.g. $V_{\mathrm{U}}/V_{\mathrm{conf}}$,
$\rho_{\mathrm{BP}}$/$\rho_{\mathrm{N}}$ and
$\rho_{\mathrm{AGN}}/\rho_{\mathrm{N}}$ 
are larger than
the values chosen in Eq.(\ref{int1}) to give a comparable
total cosmic-ray energy density.
With this alternative choice of parameter values 
Galactic SNRs would contribute only a small
fraction to the locally observed hadronic cosmic rays.
Such a situation would be in good accord with the
observational data discussed in Sect.\ref{observation}.
A value larger than 1 for $V_{\mathrm{U}} / V_{\mathrm{conf}}$
is plausible because D$_{\mathrm{IG}}$ is possibly
not large enough to allow a filling of the cosmological
``voids'' in the matter distribution (see the discussion
in Sect.\ref{diffsec}). The effective confinement volume
for the bulk of cosmic radiation below a certain energy would
then be the 
Galaxy rich ``walls'', of which the local supergalaxy
is a part. The confinement volume of intergalactic cosmic rays
was estimated to fill a fraction of about one percent of the
total volume of the universe by 
Brecher $\&$ Burbidge (\cite{brecher}).
A modern rough estimate for volume of the walls 
is about 20 to 40 $\%$ of the total volume of the universe
(assuming a thickness of
6/h Mpc for the sheets, and a size of 50 - 100 Mpc/h
for the voids, where h is the Hubble constant 
in units of 100 km/sec/Mpc (Peebles \cite{peebles})).
\\
It is presently a completely open question
if the intergalactic hadronic cosmic-ray density
was mainly produced in normal galaxies or
active objects. If the latter dominates, radio galaxies
like Cen A seem to be the most natural accelerators
for the locally observed hadronic cosmic rays
in the present scenario. The low ambient matter density
in their giant radio lobes probably allows the acceleration
of heavy elements with negligible spallation processing.
The acceleration of extragalactic cosmic rays in radio galaxies
has been discussed in detail by Rachen $\&$ Biermann
(\cite{rachen}) in the context of cosmic rays with very
high energies above 10$^{18}$ eV\footnote{
Rachen $\&$ Biermann (\cite{rachen}) 
assume that mainly protons are
accelerated in this environment. Whether or not this
assumption is true, depends on the question whether
the intergalactic medium in the surroundings of the
radio galaxy is primordial or consists of chemically processed
plasma from an early generation of active objects. As
discussed in Sect.\ref{diffsec}, it is presently not clear 
which option is correct. Especially in the local supergalaxy
the second option might be closer to reality. 
Nath $\&$ Trentham (\cite{nath}) recently argued that the
intergalactic medium was enriched with heavy elements
already at very early times. Another
possibility for extragalactical acceleration with non-primordial
abundances 
is an acceleration site in the inner (but nonnuclear)
region of active objects (Norman et al.
\cite{norman}). It is therefore quite possible that
condition 4. in Sect.\ref{flux} is fulfilled.
\label{foot8}
}.
Another plausible option is that the intergalactic cosmic-ray
energy density is mainly supplied by galaxies with a very high
rate of star formation, like 
galaxies in a starburst phase. They develop
strong galactic winds which lead to termination shocks
at which efficient particle acceleration can occur (V\"olk et al.
\cite{voelk}). 
In this case hadronic cosmic rays would be mainly produced
in {\it extragalactic} termination shocks.
\vspace{-0.15cm}
\section{Speculative extragalactic 
scenario for an explanation
of the cosmic-ray energy spectrum}
\label{scenario}
In this section I demonstrate that the ``enhancement mechanism''
of section\ref{flux} together with standard ideas about
Galactic propagation of cosmic rays is consistent
with the observed energy spectrum of cosmic rays.
%The position of the so called ``ankle'' in the energy spectrum
%at about 3 $\cdot$ 10$^{18}$ eV is explained.
The purpose of the proposed scenario is only
to demonstrate that a consistent explanation 
does not seem to meet unsurmountable difficulties.
It is virtually certain that the whole truth
is much more complicated than the simple picture
outlined below.
First I assume that the ``knee'' feature, 
where the observed
all-particle cosmic-ray spectrum 
(Berezinskii et al. \cite{berez}) steepens
from a power law with a differential index of $\alpha$=-2.6
to  $\alpha$=-3.0 at an energy of 2 PeV is already present
in the local intergalactic spectrum.
This intergalactic spectrum is assumed to
steepen from  $\alpha$=-2.1 below the knee to
$\alpha$=-2.5 above.
If hadronic cosmic rays are confined to certain regions of the
universe like the local supergalaxy (see previous
section), the ``knee'' could mark
the energy where the cosmic rays are no longer completely
confined and begin to leak out into the cosmological voids.
In this case cosmic rays of very high energy would be much
younger than the age of the universe t$_{\mathrm{U}}$.
\begin{figure}
	 \epsfxsize=8.7cm
	 \epsffile{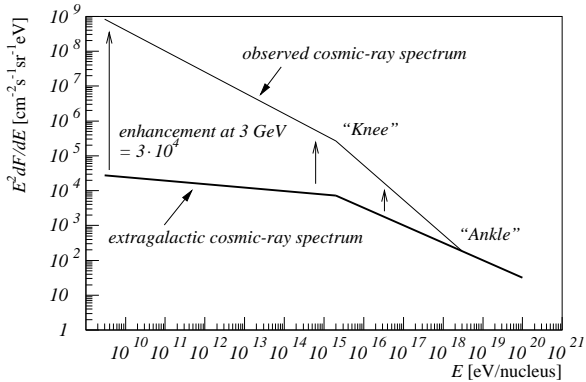}
	\vspace{-0.5cm}
         \caption[]{\small \label{originfigure3.eps}
          An extragalactical scenario for the origin 
          of the observed cosmic-ray spectrum near earth. Plotted
           is the differential flux of hadronic cosmic rays
           multiplied by E$^2$ as a function of the cosmic-ray energy
           per nucleus E . The thin line is a schematic representation
          of experimental results. They are explained by an
          extragalactical spectrum outside the Galaxy (thick line) which
           is enhanced in the Galaxy by
          a factor $e$ (symbolized by the arrows). This factor
           varies $\sim$ E$^{-0.5}$. Only at energies above the
           ``ankle'' the extragalactical spectrum is equal to the
           observed spectrum.} 
	\vspace{-0.3cm}
\end{figure}
I will further assume the following
dependence of the Galactic diffusion coefficient D$_{\mathrm{G}}$
on the cosmic-ray energy E:
\begin{equation}
D \sim E^{a}
\label{diff}
\end{equation} 
with a =0.5. This value for a is
in agreement with experimental observations
and theoretically plausible (Ferrando \cite{ferrando}).
According to Eq.(\ref{enhance}) and Eq.(\ref{diff}) I get:
\begin{equation}
e = 3 \cdot 10^{4} (E/3 GeV)^{-0.5}
\end{equation} 
This leads to an expected local spectrum in agreement
with observations (fig.1).
 $e$=1 (i.e. no more ``flux trapping'' enhancement over
the intergalactic density) is therefore expected at
an energy of

\noindent E$_{\mathrm{e=1}}$= 3 GeV $\cdot$ $e^{1/a}$  
$\simeq$ 3 $\cdot$ 10$^{18}$ eV

This is about the energy where the 
power-law index of the observed cosmic-ray
spectrum is observed to change from -3.0
to -2.6 at higher energies (``ankle'').
Detailed studies show that the observed spectrum
is described by a superposition of a steep component
at low energies and component with $\alpha$=-2.5 at
higher energies (Sokolsky \cite{sokolsky}).
An extragalactical origin of cosmic rays
above the ``ankle'' has long been considered likely.
The power law index and intensity of this component
follow quantitatively as a natural consequence
of our scenario.
\section{Observational arguments in favour of an extragalactic
origin}
\label{observation}
\vspace{-0.15cm}
After refuting an often made counter argument
for an extragalactical origin of the heavy elements
in the cosmic rays, I discuss two recent pieces of
experimental evidence which could point towards
a non Galactic origin of hadronic cosmic rays.
\\
Longair (Longair \cite{longair}) quotes as evidence
{\it against} an extragalactic origin of cosmic
rays the fact that cosmic-ray clocks like $^{10}$Be
indicate an ``age'' of cosmic rays of a few tens of million
years (Ferrando \cite{ferrando}). This is much less
than the expected time since acceleration 
in an extragalactic scenario, which is on the order of
the age of the universe t$_{\mathrm{U}}$. What cosmic rays clocks
measure, however, is the time since they 
have been propagating in a medium
dense enough to lead to
nonnegligible spallation processing (the radioactive $^{10}$Be
is a spallation product 
from nuclear reactions during propagation
and not a remnant from the acceleration site).
If extragalactic cosmic rays were accelerated in regions with
low matter density and then propagated in the intergalactic
medium which has a very low ambient density, the measured
``age'' merely measures the time since entering
the Galaxy, which is on the order of t$_{\mathrm{conf}}$, like
in the Galactic scenario of cosmic-ray origin.
\\
The first argument concerns the high-energy $\gamma$-ray
emission from Galactic SNRs.
If hadronic cosmic rays are of Galactic origin, only
SNRs seem to able to accelerate enough particles
to replenish the cosmic rays lost to intergalactic
space (Biermann \cite{biermann}). 
While the properties of the observed hadronic 
cosmic rays and the ones 
theoretically expected from SNRs
under simple assumptions
are not in perfect agreement\footnote{
E.g. Drury (\cite{drury}) noted that 
the absence of any feature in the observed
hadronic cosmic-ray spectrum around the
expected theoretical upper cut-off energy (about 100 TeV
under the simplest assumptions) is
unexpected if the observed hadronic
cosmic-rays have a SNR origin.},
a first order Fermi acceleration of
particles in these objects seems eminently
plausible(Berezhko $\&$ V\"olk \cite{berezhko}).
\\
The mean energy put into the production
of hadronic cosmic rays for a typical SNR is constrained
to be on the order of  10$^{50}$ erg by the requirement
to supply the local cosmic radiation.
VHE and UHE $\gamma$-ray emission from these objects
is then expected due to $\pi_0$ production in cosmic-ray
interactions with ambient  matter 
(e.g. molecular clouds) and has been quantitatively
calculated in several
seminal papers by a Heidelberg/Dublin group
(e.g. Aharonian et al. (\cite{aharonian})) and others
(Baring et al. \cite{baring}).
The predicted levels are on the order of the
sensitivities of various ground based
$\gamma$-ray detectors.
The nondetection of 
VHE and UHE $\gamma$-radiation from
shell type SNRs (Hess \cite{hess}; Lessard
et al. \cite{lessard}; Prosch et al.\cite{prosch}) 
is therefore becoming problematic for the Galactic
senario.
Similar detailed work
(identification of plausible sources for the locally
observed hadronic cosmic rays,
calculation of predicted $\gamma$-intensities followed by
observations) under the
assumption of an extragalactic
scenario for cosmic-ray origin is urgently needed.
\\
A second completely idenpendent argument concerns
the distribution of hadronic 
cosmic rays as a function of galactocentric radius.
Recent measurements by the EGRET experiment confirmed
the earlier conclusion drawn from data taken with 
the COS-B satellite that the galactocentric 
falling gradient of the cosmic ray 
density is much smaller than expected for
any plausible class of Galactic cosmic-ray sources
(Erlykin et al. \cite{erlykin}).
The only natural possibility to explain this within the
Galactic scenario (where the production 
of the major part of cosmic rays is expected to take 
place within the solar circle)
seems to be to assume a confinement
volume in an extended halo of the
Galaxy (extension 
from the Galactic center with a length scale of 
r$_{\mathrm{e}}$ $>$ 15 kpc)(Berezinskii et al. \cite{berez})
in which the cosmic-ray density would be roughly
constant.
This in turn would lead to the expectation of a cosmic-ray
anisotropy in the direction towards the antigalactic
center due to the Compton-Getting effect which
can be roughly estimated as:
\begin{equation}
\begin{array}{rcl}
\delta &\simeq& {(\alpha+2) (v_{\mathrm{out}}/c) } \simeq
(\alpha+2) D_{\mathrm{G}}/r_{\mathrm{e}} c  \\ &\simeq& 0.1 (E/100 TeV)^{0.5}
\end{array}
\end{equation}
Here $\alpha$ is the differential power-law index of the cosmic-ray
energy spectrum, and v$_{\mathrm{out}}$ the effective
streaming velocity away from the central
region of the Galaxy.
The energy dependence
of D$_{\mathrm{G}}$
from Eq.(\ref{diff}) was assumed. Moreover
I set D$_{\mathrm{G}}$ $\simeq$ 2 $\cdot$ 10$^{29}$ cm$^2$ sec$^{-1}$, 
in order to reproduce the empirical
storage times of smaller than a few 10$^8$ years together
with the existence of an extended halo.
The predicted value lies more than two orders
of magnitude above the observed value
of $\delta$ $\simeq$ 8$\cdot$ 10$^{-4}$
(Munakata et al. \cite{munakata})
around 100 TeV primary energy.
Though it might be possible to find a model
with very different diffusion coefficients in the
disk and halo and/or cosmic-ray transport via Galactic winds
(Erlykin et al. \cite{erlykin2})
which explains the observations within
the Galactic scenario, the magnitude of the discrepancy
seems severe.
\\
In the extragalactic picture the Compton-Getting effects
from Galactic rotation and proper motion of the Galaxy
are suppressed by a factor $e$ (Eq.(\ref{enhance}))
(Brecher $\&$ Burbidge \cite{brecher}) 
and therefore to small to be presently observable.  
To zeroth order no galactocentric gradient at all
is expected in the extragalactic scenario.
The small observed gradient (Erlykin \cite{erlykin})
can be explained as due to a contribution by
Bremsstrahlung from electrons which 
are presumably of Galactic origin (see
next section) and therefore do show a gradient.
A small gradient of the hadronic cosmic rays could be due
to a slight decrease of $e$ with galactic radius.
This might be expected if the Galactic confinement
volume is not exactly spherical or cylindrical but shaped
like a thick disk decreasing in thickness with
galactocentric radius.
\vspace{-0.2cm}
\section{The non-detection of the SMC in 
the light of $\gamma$-rays}
\label{smc}
Sreekumar et al.(\cite{sreekumar}) have argued that their
non-detection of $\gamma$-radiation above 100 MeV
from the Small Magellanic cloud (SMC) rules out an extragalactic
origin of cosmic-rays.  
Taking into account magnetic-flux trapping it is clear
that the density of cosmic rays in the SMC is not
simply expected to be equal to the local one near earth, as assumed
by Sreekumar at al..
Rather it is given by the local density times 
a factor {$e_{\mathrm{SMC}} / e$}.
Here $e_{\mathrm{SMC}}$ is the enhancement
factor valid for the SMC analogous to $e$ (Eq.(\ref{enhance})) 
for the Galaxy.
$e_{\mathrm{SMC}}$ is probably smaller than $e$ because
the corresponding t$_{\mathrm{conf}}$(SMC) is smaller than
the Galactic value
due to the smaller size and possibly dynamical
disintegrating state of the SMC.
A smaller t$_{\mathrm{conf}}$(SMC) is also directly responsible for the low
$\gamma$-ray luminosity of the SMC in the Galactic
scenario for cosmic-ray origin.
It is thus clear that the $\gamma$-ray luminosity of the SMC
is no suitable testing ground for a decision in the
question of Galactic versus extragalactic origin of cosmic rays.
\\
\vspace{-0.15cm}
\section{Conclusion}
The assumptions made for the basic hypothesis 
of flux trapping (listed
in Sect.\ref{flux}) are speculative and controversial.
A better understanding of very complex magnetohydrodynamic
processes in intergalactic space
is needed  to make
a firm decision whether they are realistic or not.
Unfortunately all present theories of cosmic-ray origin
need to assume some unproven facts. 
From a purely phenomenological point of view the
following point of view seems interesting:
\\
Galactic SNRs produce the observed 
electron cosmic-ray flux. The required 
acceleration efficiency is modest
and the high-energy cutoff lies in the region
of 10 TeV, in good agreement with theoretical
expectation 
(Mastichiades $\&$ de Jager \cite{masti}).
Protons and nuclei are perhaps accelerated with a  roughly similar
efficiency (i.e. much less than with a 100 times higher
efficiency as required in a scenario with Galactic origin) and high-energy
cutoff (see Sect.\ref{origin}), and therefore produce a
local hadronic cosmic-ray intensity similar
to the electron intensity (about 1 $\%$ of the total
intensity below the cutoff). The main part of hadronic cosmic rays
is due to intergalactic cosmic radiation which has
an enhanced density in the Galactic confinement
volume 
\footnote{
The present situation in the question of hadronic cosmic-ray
origin (HCR) has some parallels to the 
situation in the question of
the nature of $\gamma$-ray bursts (GRB) until very recently
(Paczynski \cite{paczynski}):
After a long period (HCR: 30 years,
GRBs: 15 years) in which a Galactic origin was accepted
nearly unanimously in the community, new data
(HCR: weak Galactocentric gradient in
combination with high degree of
isotropy, absence of 
VHE/UHE $\gamma$-rays from SNRs; GRB:number-luminosity
relationship in
combination with high degree of isotropy, time-dilation 
effects in bursts)
and a growing appreciation for the possibility of certain 
physical processes
(HCR: magnetic-flux trapping, GRB:  formation of relativistic fireballs)
forces us to take the ``extragalactic alternative'' serious.
In both cases only 
a localization in a very extended Galactic halo remains
a possibility in an ``undecided'' period, where both
models are taken serious.}.
\\
In the scenario of Sect.\ref{scenario}
cosmic rays with extremely high
energies have the same origin as the bulk of
hadronic cosmic rays. This gives added importance to
an experiment like Auger (Mantsch \cite{mantsch}),
which tries to find the site of origin of cosmic rays
with the highest energies. Experiments
searching for antimatter in cosmic-rays
of extragalactic origin  like AMS (Ahlen et al. \cite{ahlen}) would
have a sensitivity orders of magnitude higher than
previously thought.

\vspace{-0.15cm}
\section*{Acknowledgements}
I would like to thank D.Petry and S.Pezzoni for
useful discussions and help with the figures.
E.Lorenz and H.V\"olk helped me with critical
remarks on the manuscript. This work
was supported by a Heisenberg fellowship of
the Deutsche Forschungsgemeinschaft.

% Berezhko,E.G.,V\"olk,H.J., MPI-V12-1997, to appear in
%Astropart. Phys.
\end{document}